\documentclass[aps,pra,pdf,superscriptaddress,amsmath,amssymb,amsfonts,twocolumn,showpacs,longbibliography]{revtex4-2}

\usepackage{amssymb}
\usepackage{eqnarray,amsmath}
 
\usepackage{graphicx}
\usepackage{epsfig}
\usepackage{dcolumn}
\usepackage{bm}
\usepackage{braket}
\usepackage{amsmath}
\usepackage{physics}
\usepackage{csquotes}
\usepackage{mathtools}
\usepackage{graphicx,color,xcolor,colortbl}

\usepackage[colorlinks=true,
linkcolor=blue,
filecolor=blue,      
urlcolor=blue,
citecolor=blue]{hyperref}

\usepackage[normalem]{ulem}

\definecolor{Gray}{gray}{0.85}
\definecolor{LightCyan}{rgb}{0.88,1,1}
\definecolor{DarkGreen}{RGB}{0,100,0}

\newcolumntype{a}{>{\columncolor{Gray}}c}
\newcolumntype{b}{>{\columncolor{white}}c}

\begin{document}

\title{Dimensional crossover of bound complexes in a two-species Bose-Hubbard lattice: Correlations and dynamics}

\author{Deepak Gaur}
\affiliation{Mathematical Physics and NanoLund, LTH, Lund University, Box 118, 22100 Lund, Sweden}

\author{Koushik Mukherjee}
\affiliation{Department of Engineering Science, University of Electro-Communications, Tokyo 182-8585, Japan}
\affiliation{Mathematical Physics and NanoLund, LTH, Lund University, Box 118, 22100 Lund, Sweden}

\author{Stephanie M. Reimann}
\affiliation{Mathematical Physics and NanoLund, LTH, Lund University, Box 118, 22100 Lund, Sweden}

\begin{abstract}
We study the equilibrium and nonequilibrium formation of four-particle complexes in a balanced two-species Bose-Hubbard model with repulsive intraspecies and attractive interspecies interactions. Using exact diagonalization, we characterize the transition from weakly to strongly correlated dimer and tetramer states along the one- to two-dimensional crossover in coupled-chain geometries by combining local correlation signatures with global diagnostics such as the binding energy and interspecies entanglement entropy. We show that transverse connectivity between chains qualitatively reshapes the phase diagram, substantially enlarging the tetramer region and, in particular, stabilizing weakly bound tetramers when compared to the one-dimensional chains. By tuning the interchain hopping, we identify a transition from a degenerate manifold of spatially separated dimers to a localized tetramer ground state, driven by the lifting of one-dimensional configurational degeneracies and an associated kinetic-energy gain. By extending our analysis to larger complexes of six and eight particles, we reveal that these binding mechanisms are robust and suggest a general pathway for the formation of higher-order complexes. Finally, we demonstrate interaction and geometric quench protocols to dynamically prepare these complexes with high fidelity. Our results provide a microscopic framework for engineering and probing few-body bosonic bound states in tunable lattice geometries.
\end{abstract}

\maketitle


\section{Introduction}

Ultracold atoms provide an exceptionally clean and tunable platform to explore few-body bound complexes~\cite{bulgac_2002,mistakidis_2023}, with independent control over interactions, dimensionality, and geometry enabled by Feshbach resonances and optical lattices~\cite{chin_2010,greiner_2002,bloch_2005,morsch_2006,lewenstein_2007,bloch_2008}.
This versatility has fueled intense interest in self-bound quantum matter, including quantum droplets and other related phases, on both the theoretical~\cite{bulgac_2002,petrov_2015,petrov_2016,astrakharchik_2018,mukherjee_2023,brauneis_2025,englezos_2025} and experimental fronts~\cite{kadau2016,igor_2016,schmitt_2016,chomaz_2016, cabrera_2018,cheiney_2018,igor_2018,semighini_2018,houwman_2024,cavicchioli_2025}. 
Few-body precursors of bound states ~\cite{bulgac_2002} were particularly discussed also for low-dimensional geometries; see, for example, Refs.~\cite{tempfli_2009,bougas_2021,chergui_2023,mistakidis_2023,friethjof_2024,chergui_2025,backert_2025}. 
In binary mixtures, attractive interspecies interactions may lead to the formation of composite dimers, which can further cluster into higher-order bound states such as tetramers or hexamers~\cite{guijarro_2020}. These few-body complexes form the fundamental building blocks of larger self-bound structures; for instance, in mass-balanced one-dimensional (1D) Bose-Bose mixtures, a crossover from repulsive to attractive effective dimer-dimer interactions is predicted to signal the emergence of a dilute dimerized liquid~\cite{pricoupenko_2018}.

Optical lattices add a further layer of complexity by imposing spatial discreteness, enabling Hubbard-type descriptions~\cite{hubbard_1963,fisher_1989,jaksch_1998,greiner_2002,bloch_2005,morsch_2006,lewenstein_2007,bloch_2008} and facilitating the emergence of exotic bound states that lack a continuum counterpart, such as repulsively bound pairs~\cite{winkler_2006}. Theoretical exploration has revealed a rich hierarchy of lattice-bound complexes, including dimers, trimers, and multimers existing outside the standard scattering continuum~\cite{petrosyan_2010,valiente_2008}, as well as lattice-induced Efimov-like states~\cite{nygaard_2008,valiente_2010}. For 1D mixtures, the predicted formation of dimerized liquids has been characterized using density-matrix renormalization group techniques and effective hard-core dimer models~\cite{morera_2020,morera_2021,jofre_2024}. Recently, few-body bound clusters in small two-dimensional (2D) lattices have been studied using exact diagonalization~\cite{matias_2025}.

Despite significant advances in lattice pairing and effective dimer descriptions, the microscopic mechanisms governing tetramer formation in dilute two-component bosonic lattices, and their dependence on geometry and dimensional crossover, remain largely incomplete, especially in a non-equilibrium setting. 
In particular, there is a need for (1) a systematic mapping of few-body complexes across interaction parameters using diagnostics that go beyond energy considerations alone, (2) a reliable discrimination between weakly bound and strongly bound four-body states in higher-dimensional lattices, including their characteristic correlation signatures, and (3) dynamical protocols demonstrating how these complexes can be prepared and interconverted in real time. 
Addressing these challenges is timely in view of recent experimental advances enabling the deterministic preparation and site-resolved detection of few-particle systems in microtraps~\cite{serwane_2011,wenz_2013,murmann_2016, bayha_2020}, optical tweezer arrays~\cite{kaufman_2021,spar_2022}, and optical lattices equipped with the capability of quantum gas microscopes~\cite{bakr_2009,sherson_2010,weitenberg_2011,cheuk_2015,omran_2015,haller_2015,parsons_2015,gross_2017,gross_2021,tao_2024} to directly image real-space correlations with single-site resolution.

In this work, we investigate a particle- and mass-balanced two-species Bose-Hubbard system with repulsive intraspecies and attractive interspecies interactions between bosons of kinds $A$ and $B$. In such a mixture, 
particles generally tend to maximize their binding by forming 
the largest possible bound cluster allowed by the total particle count $N$. 
In the following, we primarily focus on the $N_A = N_B = 2$ case where this maximum cluster corresponds to a tetramer. Using exact diagonalization in real space, we first establish a comprehensive equilibrium picture of few-body complexes in a strictly 1D chain and in coupled-chain geometries that interpolate continuously toward the 2D limit. 
This minimal setting of four particles, enforced by the computational restriction with the exact-diagonalization methodology employed in our work, serves the purpose of elucidating lattice-mediated stability of few-body bound complexes.
Due to the attractive interspecies interaction, the ground state always corresponds to a dimerized state, with a local pairing between $A$ and $B$ species. Note that this dimerization vanishes and the ground state corresponds to a four-particle scattering state when the attractive interaction is switched off. We map the occurrence of weakly and strongly correlated \footnote{By correlation, in the context of ``weakly and strongly correlated states'', we mean interspecies correlation which causes the pairing between particles of different species and binds them into two-body and four-body clusters.} 
dimer and tetramer states across the interaction parameter space using a set of complementary diagnostics: the four-body binding energy, the interspecies von Neumann entanglement entropy (including its curvature in the parameter space to locate the weakly correlated -- strongly correlated dimer crossover), and spatially resolved one- and two-body density matrices that directly expose pairing and clustering in real space. This approach provides a consistent microscopic classification of weakly correlated dimerized states, strongly correlated dimer pairs, and four-body bound complexes, and it enables us to resolve distinct correlation fingerprints beyond energy-based criteria alone.

A central result of our study is that lattice geometry and transverse connectivity between 1D chains reshape the stability and internal structure of tetramers. By systematically increasing the number of coupled chains and by independently tuning the interchain tunneling $J_y$, we show that the tetramer region expands substantially upon approaching a spatially symmetric 2D configuration. Most importantly, we identify a narrow but robust parameter corridor in which tetramers persist even when the constituent $AB$ pairs are only weakly bound and spatially extended for large intraspecies repulsion. This regime is absent in strictly 1D chains, and the underlying bound-state structure can be distinguished from strongly bound tetramers through characteristic two-body correlation signatures.
We further demonstrate that varying $J_y$ drives a geometry-controlled reordering of the ground-state manifold: A highly degenerate set of spatially separated dimers in the decoupled limit evolves into a localized tetramer ground state as transverse
hopping lifts 1D configurational degeneracies and provides a kinetic-energy gain. While the main analysis focuses on the $N_A = N_B = 2$ system to maintain numerical exactness across various lattice geometries, we also examine larger particle numbers and identify higher-order bound complexes, such as hexamers and octamers. These results establish that the role of dimensionality in stabilizing bound states is not restricted to few-body tetramers, but persists for larger composite structures.

Finally, guided by the equilibrium phase diagrams, we establish nonequilibrium protocols, such as interaction ramps and geometric quenches, that dynamically prepare and interconvert dimer and tetramer states with high fidelity, while tracking the
real-time buildup of pairing and clustering via density matrices, binding energy, and entanglement growth. Together, these results provide a microscopic route to engineer and diagnose few-body bound complexes in tunable lattices that are directly compatible with current capabilities for deterministic few-particle preparation and site-resolved detection.

The remainder of this article is organized as follows. Section~\ref{sec_theory} introduces the model Hamiltonian, numerical methods, and key observables. Section~\ref{Eql_phaes} presents the equilibrium phase diagrams and correlation signatures of $N_A = N_B= 2$ system for both strictly 1D lattices (Sec.~\ref{only_1D}) and coupled arrays of 1D lattices. The latter explores two geometric controls: varying the number of chains (Sec.~\ref{tune_ly}) and tuning the interchain hopping $J_y$	(Sec.~\ref{tune_jy}). In Sec.~\ref{diff_particles}, we investigate the stability of higher-order complexes by extending our analysis to larger particle numbers ($N=6$ and $N=8$).
In Sec.~\ref{Dynamics}, we investigate the nonequilibrium formation of few-body bound states, focusing on interaction quenches (Sec.~\ref{sec_dyn_intquench}) and on dimensional crossovers induced by quenching the interchain hopping strength (Sec.~\ref{dimen_cross}). Finally, in Sec.~\ref{sec_conclusions}, we summarize our main findings and outline possible future directions. Additional insights into the formation of two-particle bound state ($AB$ pair) are provided in Appendix~\ref{appendA}. In Appendix~\ref{appendB}, we provide further discussion on the effects of open and periodic boundary conditions.

\section{Theoretical Model} \label{sec_theory}

\subsection{Setup and Hamiltonian}

We consider a Bose-Bose mixture interacting via only short-range contact interactions and confined in an optical lattice, which could be either one- or two-dimensional. The mixture can correspond, for example, to two different hyperfine states of the same atomic species. The system is modeled by the two-species Bose-Hubbard Hamiltonian, which reads~\cite{damski_2003}

\begin{eqnarray}
  \hat{H} &=& \sum_{i,j,\sigma} \Big[
           - J_x \, \hat{b}^{\dagger\sigma}_{i+1,j}\hat{b}^{\sigma}_{i,j}
           - J_y \, \hat{b}^{\dagger\sigma}_{i,j+1}\hat{b}^{\sigma}_{i,j}
           + \mathrm{H.c.} \Big]
   \nonumber\\
          &&+ \sum_{i,j} \bigg[ \sum_{\sigma} \frac{U}{2} \,
           \hat{n}^{\sigma}_{i,j} \big(\hat{n}^{\sigma}_{i,j}-1\big)
           + U_{AB} \, \hat{n}^{A}_{i,j}\hat{n}^{B}_{i,j} \bigg].
 \label{ham}
\end{eqnarray}

Here, $\sigma = (A,B)$ labels the two bosonic species, and $(i,j)$ denotes the lattice site indices of a 2D square lattice. $J_x$ ($J_y$) is the nearest-neighbor hopping amplitude along the $x$ ($y$) direction for both species. The operators $\hat{b}^{\sigma}_{i,j}$ and $\hat{b}^{\dagger \sigma}_{i,j}$ annihilate and create, respectively, a boson of species $\sigma$ at site $(i,j)$, and
$\hat{n}^{\sigma}_{i,j} = \hat{b}^{\dagger \sigma}_{i,j}\hat{b}^{\sigma}_{i,j}$ is the corresponding number operator. In the 1D limit, the Hamiltonian reduces to the corresponding 1D form upon setting $J_y=0$ and replacing the site index $(i,j)$ with $i$.

Throughout this work, we consider positive hopping amplitudes $J_x, J_y \geq 0$ and repulsive intraspecies on-site interactions $U > 0$, both taken to be species independent. The interspecies on-site interaction is chosen to be attractive, $U_{AB} < 0$. Finally, all energy scales are measured in units of the hopping amplitude $J_x$, which is set to unity ($J_x = 1$).
It should be noted that the parameters of the model Hamiltonian can be expressed in terms of the recoil energy ($E_R$) and are ultimately dependent on the lattice depth, lattice constant, $s$-wave scattering length, and Wannier functions of the lowest band. We refer to refs.~\cite{jaksch_1998,blakie_2004}, which provide an estimation of these parameters using the microscopic Hamiltonian.

\subsection{The many-body wavefunction and method}

To calculate the stationary and the nonequilibrium dynamical properties of the system described above, we employ the exact-diagonalization method in real space. In this method, various choices for the distribution of particles of the two species constitute the basis states of the system. The single-species basis states on an $L_x \times L_y$ lattice are denoted as $\ket{\phi_k^{\sigma}} \equiv \ket{n^{\sigma}_{0,0}, n^{\sigma}_{0,1}, \cdots n^{\sigma}_{L_x-1,L_y-1} }_k$, where $n^{\sigma}_{i,j}$ represents the occupancy of species $\sigma$ at site $(i,j)$ for the $k$th possible configuration of particle distribution. 
We remark that $L_x$ ($L_y$) denotes the number of lattice sites along $x$ ($y$) direction of the 2D square lattice and refers to the size of the lattice in units of the lattice constant $a$. Also, in our notation, lattice site indices start from 0.
The Hilbert space of the full system then consists of basis states of the form $\ket{\phi^A_{k}} \otimes \ket{\phi^B_{l}}$. The many-body wavefunction of the system can thus be expressed as
\begin{equation}
    \ket{\Psi} = \sum_{k, l} C_{kl} \ket{\phi^A_{k}} \otimes \ket{\phi^B_{l}} 
    \equiv \sum_m C_m \ket{\psi_m},
    \label{many_bdy_wave}
\end{equation}
where, $\ket{\psi_m}\equiv \ket{\phi^A_{k}} \otimes \ket{\phi^B_{l}}$ are basis states corresponding to the various configurations of particles of the two species $A$ and $B$ on the lattice.

In this work, we use coupled basis states $\ket{\psi_m}$ to construct the Hamiltonian matrix $H_{m m'} = \bra{\psi_{m}} \hat{H} \ket{\psi_{m'}}$, with open boundary conditions (OBCs). We have also verified that the observed phases are robust against finite-size effects. A detailed comparison between OBCs and periodic boundary conditions (PBCs) is provided in Appendix~\ref{appendB}, confirming that the qualitative physics remains unchanged. The matrix is highly sparse due to short-range hopping allowed in the tight-binding model. Subsequently, we obtain the ground state and a few low-lying excited states of the system by diagonalizing the Hamiltonian matrix using the PARPACK library \cite{maschhoff_1996}. For additional details on the numerical exact-diagonalization method used in the present work, see Refs.\cite{bai_2018,gaur_2024}.

The dynamical evolution of the many-body state [Eq.~(\ref{many_bdy_wave})] is studied by performing the unitary evolution of the initial state under the time-dependent Hamiltonian as $i \partial_t \ket{\Psi (t)} = \hat{H}(t) \ket{\Psi (t)}$. Choosing $\ket{\psi_m}$ as time independent, we get
\begin{equation}
    i\frac{\partial C_{m} (t)}{\partial t} = 
       \sum_{m'}  \bra{\psi_{m}} \hat{H}(t) \ket{\psi_{m'}} C_{m'}(t).
\end{equation}
This equation represents a set of coupled partial differential equations involving time-dependent coefficients of the wavefunction $C_m (t)$. We solve these equations numerically using a fourth-order Runge-Kutta method with time step $\Delta t = 0.001$. 

\subsection{Observables}
Below, we introduce the key observables employed to characterize both the stationary and dynamical features in this work. Depending on the interspecies and intraspecies interaction strengths, the system can exhibit different states from weakly correlated to strongly correlated dimers, and bound states (see the discussion in subsequent sections). At the single-particle level, these states can be compared by examining the one-body density distribution, defined at each lattice site $(i,j)$ as
\begin{equation}
   \rho^{(1)}_{(i,j);\sigma}
      = \langle \psi | \hat{b}^{\dagger \sigma}_{i,j}\hat{b}^{\sigma}_{i,j} | \psi \rangle .
\end{equation}

To elucidate the role of intraspecies and interspecies correlations, we further consider the diagonal elements of the two-body reduced density matrix, defined as
\begin{equation}
   \rho^{(2)}_{(i,j),(i',j');\,\sigma\sigma'} =
   \langle \psi |
   \hat{b}^{\dagger \sigma'}_{i',j'} \hat{b}^{\dagger \sigma}_{i,j}
   \hat{b}^{\sigma}_{i,j} \hat{b}^{\sigma'}_{i',j'}
   | \psi \rangle .
    \label{def_2body_corr}
\end{equation}
This quantity represents the probability of simultaneously finding a boson of species $\sigma$ at site $(i,j)$ and a boson of species $\sigma'$ at site $(i',j')$. It characterizes intraspecies correlations for $\sigma'=\sigma$ and interspecies correlations for $\sigma'=\bar{\sigma}$, where $\bar{\sigma}$ denotes the other species. It thus provides direct insight into the spatially resolved localization properties of two particles relative to each other and serves as a clear indicator of the formation of different few-body complexes across interaction regimes.

The existence of few-body bound states can be further quantified by examining the binding energy, which is defined as the difference between the ground-state energy of the four-particle system, $E_{AABB}$, and the ground-state energy of two uncorrelated dimers, each with energy $E_{AB}$ (note that here a trimer is not an energetically favored state),
\begin{equation}
   E_b = E_{AABB} - 2 E_{AB}.
 \label{bind_ener_eqn}
\end{equation}
A negative value of $E_b$ signals the formation of a stable bound state.

To further distinguish the weakly correlated state from strongly correlated states, we analyze the entanglement entropy. The many-body wavefunction of the system in Eq.~(\ref{many_bdy_wave}) can be expressed according to a truncated Schmidt decomposition of rank $D$, namely,
$\ket{\Psi} = \sum_{k = 1}^{D} \sqrt{\lambda_k} \ket{\tilde{\phi}^A_{k}} \otimes \ket{\tilde{\phi}^B_{k}}$, where, $\lambda_k$ are the Schmidt coefficients \cite{ekert_1995}. 
This is achieved by a singular value decomposition of the coefficient matrix $C_{kl}= (U \lambda V)_{kl}$, where $\lambda$ is a diagonal matrix containing the Schmidt coefficients, and the matrices $U, V$ relate the rotated basis $\{\ket{\tilde{\phi}^{\sigma}}\}$ with $\{\ket{\phi^{\sigma}}\}$.
The Schmidt coefficients enter as the eigenvalues of the reduced density matrix of any one of the species (say $\rho_A$) and encode the entanglement between the two species. 
Thus, the entanglement can be quantified as the von Neumann entropy of $\rho_A$~\cite{luigi_2008} defined as
\begin{equation}
   S_{\rm N} = -{\rm Tr} \rho_A \ln{\rho_A}.
\end{equation}
The entanglement entropy provides a complementary measure of quantum correlations between the two species that constitute the system \cite{richaud_2019, richaud_2020}.  Note that for a maximally entangled state, $S_{\rm N}^{\,\max} = \ln(\min(\mathcal{D}_A, \mathcal{D}_B))$, where $\mathcal{D}_{\sigma}$ is the size of the Hilbert space of species $\sigma$.

Finally, to assess the adiabaticity of the quench dynamics and quantify the fidelity of the dynamically prepared states, we compute the many-body overlap
\begin{equation}
   \mathcal{O}(t) = \langle \Psi^{\rm eq}_{\rm gs} (H_{\rm fin}) \mid \Psi (t) \rangle,
 \label{olap}
\end{equation}
which measures the projection of the time-evolved state $\ket{\Psi(t)}$ onto the equilibrium ground state $\ket{\Psi_{\rm gs}^{\rm eq}(H_{\rm fin})}$ of the Hamiltonian at the final interaction strength.
Physically, $\mathcal{O}(t)$ provides a direct diagnostic of how closely the quench dynamics follow the ground-state manifold of the target state. Values of $\mathcal{O}(t_{\rm fin})$ approaching unity indicate near-adiabatic evolution, implying minimal excitation into higher-energy states and confirming that the final state faithfully realizes the target dimer or tetramer phase. This global observable complements the local information obtained from one- and two-body density correlations, ensuring that the observed spatial structures correspond to the intended many-body ground states.



\begin{figure*}[t]
   \includegraphics[width=0.98\textwidth]{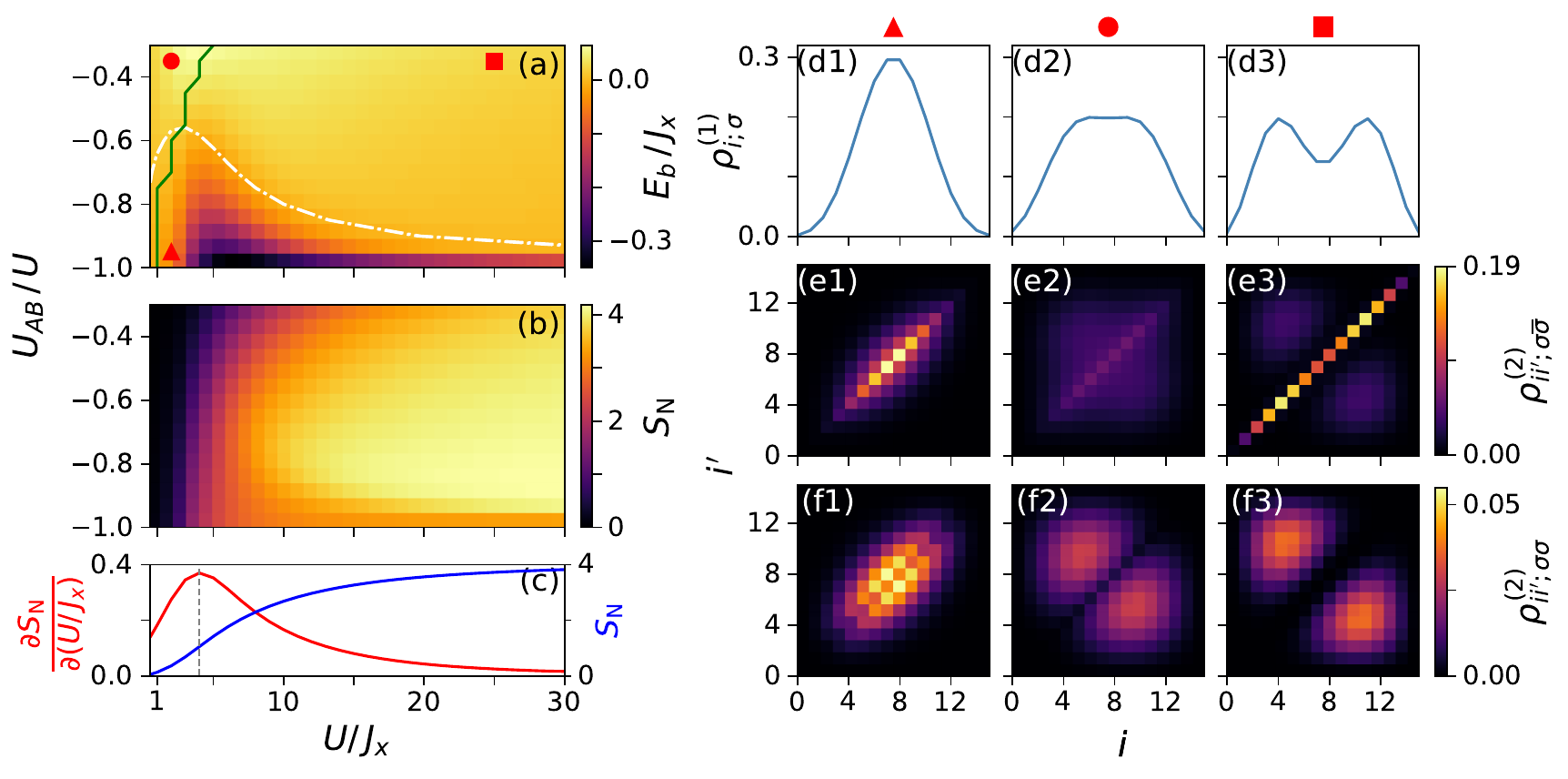}
   \caption{Ground-state few-body complexes and correlation signatures in a 1D lattice.
   (a) Binding energy, $E_b$, and (b) von Neumann entanglement entropy, $S_{\rm N}$, of the four-particle system as functions of $U_{AB}/U$ and $U/J_x$. (c) $S_{\rm N}$ (blue curve) and its derivative $\partial S_{\rm N}/\partial (U/J_x)$ (red curve) as a function of $U/J_x$ for a fixed ratio $U_{AB}/U = -0.35$.
   Panel (a) delineates the regions corresponding to the weakly and strongly correlated dimer and tetramer states. The white dash-dotted line indicates the transition to the regime of negative binding energy, signaling the formation of a tetramer. The green line demarcates the low- and high-interspecies-entanglement regimes ($S_{\mathrm{N}}^{\,\max} \approx 4.91$, for a maximally entangled state), and is identified from the peak of $\partial S_{\rm N}/\partial (U/J_x)$ [see, e.g., panel (c)].
   (d1)-(d3) One-body density distributions of species $\sigma$, $\rho^{(1)}_{i;\sigma}$, for the tetramer [panel (d1)], weakly correlated dimer [panel (d2)], and strongly correlated dimer [panel (d3)], states, respectively. Markers at the top indicate the corresponding parameter values in panel (a).
   (e1)-(e3) Interspecies two-body densities $\rho^{(2)}_{ii',\sigma\overline{\sigma}}$ and (f1)-(f3) intraspecies two-body densities $\rho^{(2)}_{ii',\sigma\sigma}$, corresponding to the same parameter points marked in panel (a).
}
   \label{pdiag1d}
\end{figure*}

\section{Equilibrium phases and correlations}\label{Eql_phaes}

In this section, we analyze the equilibrium phases and correlation properties to identify few-body bound complexes, specifically dimers and tetramers, across different interaction regimes. We first characterize these states in a single 1D lattice chain and then extend the analysis to arrays of coupled chains. This approach allows us to elucidate the role of effective dimensionality and to assess how tunable interchain hopping influences the formation and stability of these composite structures.

\subsection{1D lattice}\label{only_1D}

We begin by considering a 1D lattice of $L_x =16$ sites with open boundary conditions.  The resulting phase diagram is presented in Fig.~\ref{pdiag1d}(a), where we map the system's behavior by calculating the binding energy $E_b$ in the $U_{AB}/U$ versus $U/J_x$ plane. The tetramer region is located underneath the white dash-dotted line in Fig.~\ref{pdiag1d}(a), where $E_b < 0$ holds. As shown in Fig.~\ref{pdiag1d}(d1) (triangular marker), the tetramer is characterized by a highly localized one-body density, $\rho^{(1)}_{i,\sigma}$. 
As $U/J_x$ increases, the tetramer region in the phase diagram progressively contracts, and in the strong intraspecies interaction limit, it exists only as a narrow corridor around $U_{AB}/U \approx -1$ [see Fig.~\ref{pdiag1d}(a)].

To distinguish between the weakly and strongly correlated dimer phases outside the tetramer region, we analyze the interspecies entanglement entropy, $S_{\mathrm{N}}$ [Fig.~\ref{pdiag1d}(b)]. 
\textit{Weakly correlated dimer phase}: This phase is characterized by a low entanglement entropy, $S_{\mathrm{N}}$, and a positive binding energy ($E_b > 0$). It occupies the low-$U/J_x$ regime and is bounded by the solid green line and dashed white line in the top-left corner of Fig.~\ref{pdiag1d}(a), typically existing in the weak-interaction regime. The corresponding one-body density profile is relatively delocalized [Fig.~\ref{pdiag1d}(d2), circular marker].
\textit{Strongly correlated dimer phase}: 
As $U/J_x$ increases, the system undergoes a crossover into a strongly correlated dimer state. Although the binding energy remains positive ($E_b > 0$), this phase is distinguished by a significant increase in $S_{\mathrm{N}}$. 
The associated one-body density, $\rho^{(1)}_{i,\sigma}$, exhibits two well-separated spatial peaks, corresponding to the center-of-mass positions of two distinct dimers [Fig.~\ref{pdiag1d}(d3)] in our 1D chain.

Given that the entanglement entropy varies smoothly, we pinpoint the crossover boundary by examining the curvature of $S_{\mathrm{N}}$. Specifically, we analyze the derivative $\partial S_{\mathrm{N}} / \partial (U/J_x)$, which exhibits a characteristic concave structure [Fig.~\ref{pdiag1d}(c)]. We define the crossover point as the maximum of this derivative. The region where $\partial^{2} S_{\mathrm{N}} / \partial (U/J_x)^{2} > 0$ (to the left of the maximum) corresponds to the weakly entangled dimer phase, while $\partial^{2} S_{\mathrm{N}} / \partial (U/J_x)^{2} < 0$ (to the right of the maximum) identifies the strongly entangled dimer phase. 
Note that for a maximally entangled state, $S_{\mathrm{N}}^{\,\max} = \ln \binom{L_x + N_{\sigma} -1}{N_{\sigma}} \approx 4.91$. The classification scheme for various phases is summarized in Table~\ref{tab:summary}.

\begin{table}[t]
 \centering
  \begin{tabular}{|c|c|c|}
   \hline\hline
    $E_b$ & $\partial^2 S_{\mathrm{N}}/\partial (U/J_x)^2$ & Few-body state \\ \hline
    $E_b > 0$ & $> 0$ & Weakly correlated dimer  \\
    $E_b > 0$ & $< 0$ & Strongly correlated dimer \\
    $E_b < 0$ & $> 0$ & Weakly correlated tetramer \\
    $E_b < 0$ & $< 0$ & Strongly correlated tetramer \\
   \hline\hline
  \end{tabular}
 \caption{Classification of few-body states based on the binding energy $E_b$ and the curvature of the interspecies entanglement entropy $\partial^2 S_{\mathrm{N}}/\partial (U/J_x)^2$.}
 \label{tab:summary}
\end{table}

To provide deeper insight into the spatial structure of these states, we present the intraspecies and interspecies two-body density matrices, $\rho^{(2)}_{ii';\sigma\sigma'}$ [see Eq.~\eqref{def_2body_corr}], for the representative markers shown in Fig.~\ref{pdiag1d}(a). A maximum interspecies correlation occurs along its diagonal due to the attractive interspecies interaction [see Figs.~\ref{pdiag1d}(e1)–\ref{pdiag1d}(e3)]. In the tetramer state, this diagonal weight is sharply peaked at the center, confirming a tightly bound four-body cluster. In contrast, for the strongly and weakly correlated dimer states, the diagonal feature is more delocalized, reflecting weaker binding and increased spatial separation between particles. Turning to the intraspecies two-body density matrix $\rho^{(2)}_{ii';\sigma\sigma}$ [Figs.~\ref{pdiag1d}(f1)–\ref{pdiag1d}(f3)], we observe a depletion along the diagonal arising from the repulsive interaction $U$. This depletion is weakest for the tetramer bound state, as expected, and strongest for the strongly correlated dimer state, where the emergence of two symmetric antidiagonal lobes indicates that the two particles of the same species are spatially separated into different dimers. Specifically, Fig.~\ref{pdiag1d}(f1) reveals that particles of different species preferentially occupy nearest-neighbor sites in the tetramer. This observation implies a nearest-neighbor dimer–dimer attraction, corroborating the effective dimer model previously reported for 1D chains \cite{morera_2021}. As we demonstrate below, this behavior persists for arrays of 1D lattices.

\subsection{Array of 1D lattices}\label{Array1D}

\begin{figure}[t]
\includegraphics[width=0.48\textwidth]{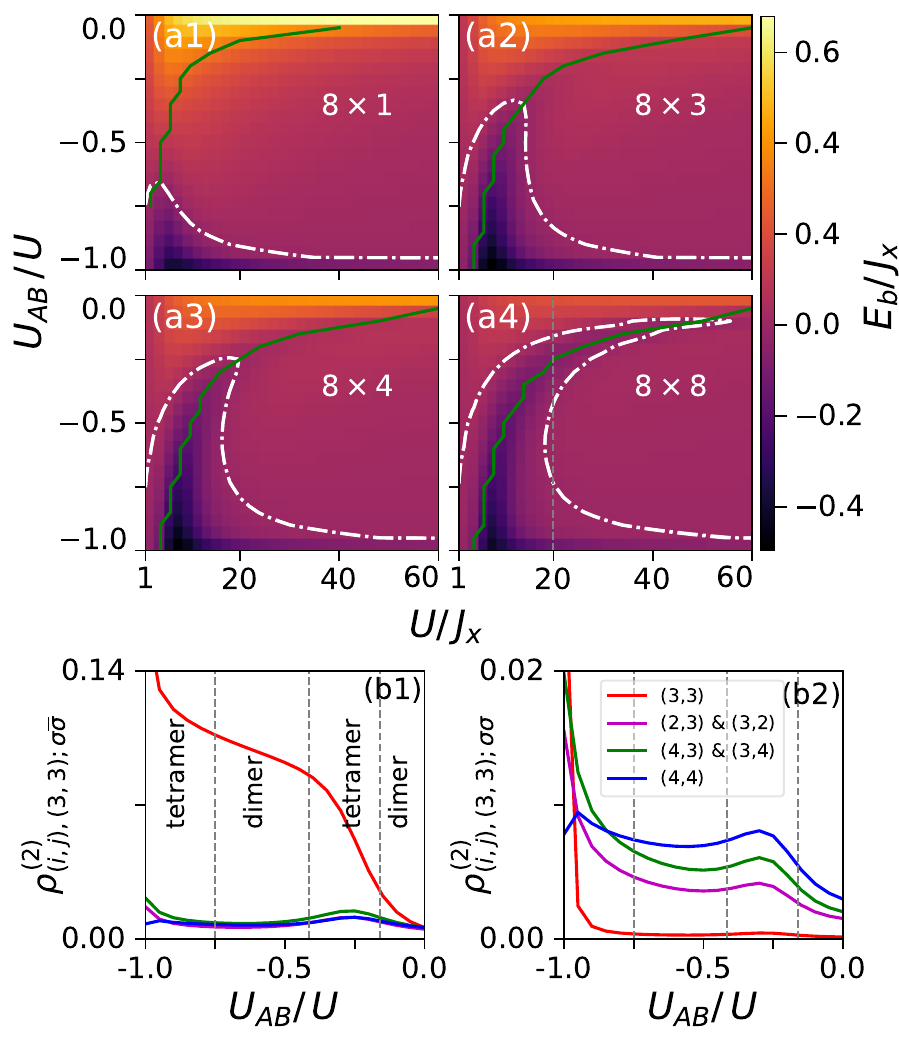}
\caption{Effect of lattice geometry. Binding energy $E_b$ as a function of $U_{AB}/U$ and $U/J_x$ for (a1) a 1D $L_x \times L_y = 8 \times 1$ lattice and (a2)–(a4) arrays of 1D lattice configurations with different $L_y$ (see legends), assuming isotropic hopping $J_y = J_x$. The tetramer phase exists over an increasingly larger region of the phase diagram as the $L_y$ increases.  For large $U/J_x$, the tetramer eventually appears in two narrow strips of the phase diagram, located around $U_{AB}/U \approx -1$ and $\approx -0.1$. The former region consists of bound states formed from strongly correlated $AB$ pairs, while the latter corresponds to weakly correlated $AB$ pairs (see text). Shown also are the behaviors of the (b1) interspecies and (b2) intraspecies correlators between a reference site  $(i', j') = (3, 3)$ in the bulk and its few neighboring sites as functions of $U_{AB}/U$ for an $8\times 8$ system at $U/J_x=20$ [dashed line in panel (a4)].}
  \label{pdiag2d}
\end{figure}

Next, we investigate how the system's phase diagram is modified by varying the lattice geometry. Specifically, we consider 
an $8 \times L_y$ geometry, where $L_y \in \{1,2,\ldots,8\}$ denotes the number of coupled 1D chains. To maintain computational feasibility while exploring the $y$ direction, we reduce the longitudinal extent from $L_x = 16$ sites to $L_x = 8$. This choice balances the exponential growth of the Hilbert space inherent to our exact-diagonalization calculations. Given the ultra-dilute nature of the system and the strictly on-site interaction model, finite-size effects remain minimal and do not qualitatively alter the underlying few-body physics. This is evidenced by the consistent binding-energy profiles obtained for the $L_x = 16$ and $L_x = 8$ in the 1D cases [compare Figs.~\ref{pdiag2d}(a1) and \ref{pdiag1d}(a)]. To this end, we examine the crossover toward the 2D limit from two complementary perspectives: 
(1) the discrete addition of 1D chains and 
(2) the continuous tuning of the interchain hopping, $J_y$.

\subsubsection{Sequential row addition ($L_y$ scaling)}\label{tune_ly} 
 The evolution of the binding energy for isotropic hopping ($J_x = J_y$) is shown in Figs.~\ref{pdiag2d}(a2)–\ref{pdiag2d}(a4). Increasing the number of rows from $L_y=1$ to $L_y=8$ results in a substantial expansion of the tetramer region ($E_b < 0$) within the $U_{AB}/U$ versus $U/J_x$ parameter space, with the phase penetrating deeper into the low-$\lvert U_{AB}/U \rvert$ regime for $L_y \gtrsim 4$ [Figs.~\ref{pdiag2d}(a3) and \ref{pdiag2d}(a4)]. Concurrently, for $U_{AB}/U > -0.5$, a sharp bending of the dimer-tetramer boundary toward higher $U/J_x$ is observed for the symmetric $8 \times 8$ lattice.  Notably, our results show that the tip of this curvature lies approximately on the green solid curve, which marks the crossover from the weakly to the strongly correlated regime.

\begin{figure*}[t]
   \includegraphics[width=0.98\textwidth]{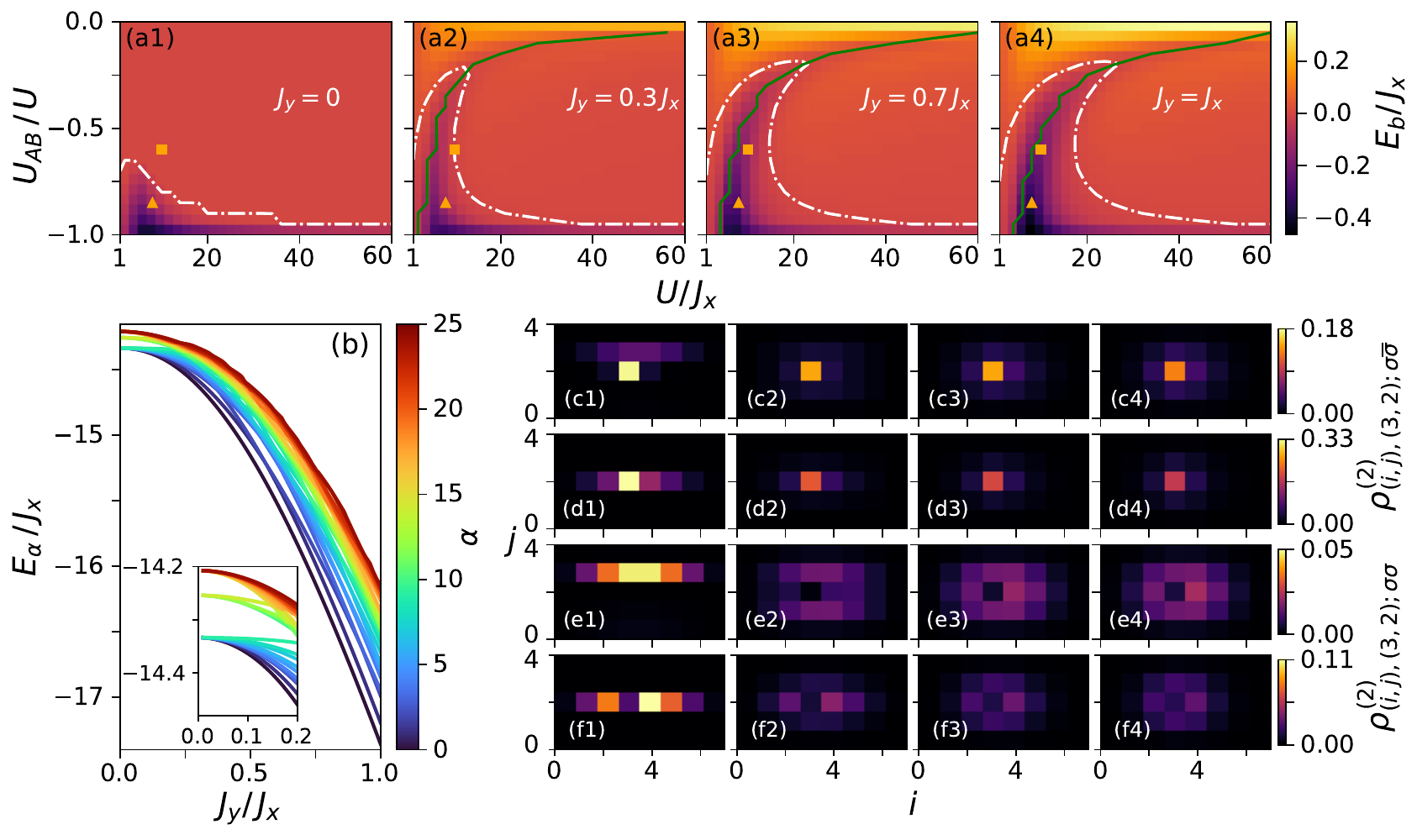}
\caption{Effect of anisotropic hopping.
(a1)–(a4) Binding energy $E_{b}$ as a function of $U_{AB}/U$ and $U/J_x$ on an $L_x \times L_y = 8 \times 5$ lattice for anisotropic hopping parameters with  $J_y/J_x = 0$ [panel (a1)], $0.3$ [panel (a2)], $0.7$ [panel (a3)], and $1$ [panel (a4)]. 
(b) Energies of the lowest 25 eigenstates as a function of $J_y$ on an $8 \times 5$ geometry for $U_{AB}/U = -0.6$ and $U/J_x = 10$. For uncoupled chains ($J_y = 0$), the chosen parameters correspond to a dimer ground state with a ten-fold degeneracy, while for $J_y = J_x$ the ground state is a tetramer. The inset further highlights the degeneracy of the ground states at $J_y = 0$. 
Two-body (c1)–(c4), (d1)–(d4) interspecies density matrices $\rho^{(2)}_{(ij),(3,2);\sigma\bar{\sigma}}$ and (e1)–(e4), (f1)-(f4) intraspecies density matrices $\rho^{(2)}_{(ij),(3,2);\sigma\sigma}$, where one particle is fixed at $(i', j') = (3, 2)$ for selected points indicated by the markers in panels (a1)–(a4). For $U_{AB}/U = -0.6$ and $U/J_x=10$, ground state switches from a dimer to a tetramer as $J_y$ increases [panels (c1)-(c4) and (e1)-(e4)] while for $U_{AB}/U = -0.85$, $U/J_x=8$ ($E_b < 0$), the ground state remains a tetramer [panels (d1)-(d4) and (f1)-(f4)].
}
  \label{pdiag_anisotropy}
\end{figure*}

Recall that the formation of the tetramer bound state originates from the spatial proximity of particles of species $A$ and $B$, driven by attractive interspecies interactions, which concomitantly lower the total energy of the system. The expansion of the tetramer region with increasing $L_y$ originates from the progressive relaxation of 1D kinematic constraints. In the strictly 1D limit, particles are restricted to a linear arrangement, which incurs a significant kinetic-energy penalty to maintain the spatial proximity required for four-body binding. As $L_y$ increases, the system gains configurational degrees of freedom along the transverse direction. 
 This occurs due to an increase in coordination number and additional transverse hopping, which allows rearrangements that reduce unfavorable on-site intraspecies occupancy.
This allows the particles to optimize attractive interspecies correlations while simultaneously minimizing the kinetic-energy cost associated with 1D confinement.

This kinetic relaxation is maximized in the symmetric 2D lattice. Consequently, even at small $|U_{AB}/U|$ \footnote{For smaller $|U_{AB}/U|$, the attractive interaction $U_{AB}/J_x \equiv (U_{AB}/U)(U/J_x)$ remains sufficiently large only when $U/J_x$ is large.}, 
the system can reach a state where all four particles remain in sufficient proximity to be weakly bound, resulting in $E_b < 0$. To further elucidate this behavior, let us concentrate on a fixed value of $U/J_x = 20$ in Fig.~\ref{pdiag2d}(a4) and analyze the evolution of interspecies and intraspecies correlators,  relative to a reference site $(i', j') = (3, 3)$, as a function of $U_{AB}/U$; see Figs.~\ref{pdiag2d}(b1) and \ref{pdiag2d}(b2). 

For $ -0.42 \lesssim U_{AB}/U \lesssim -0.16$, where $E_b < 0$, the particles within each $AB$ pair are relatively weakly bound. In this regime, the particles are distributed over a larger spatial extent, which is reflected by the sharp drop in the on-site interspecies correlator [red curve in Fig.~\ref{pdiag2d}(b1)]. This spatial delocalization is further confirmed by the relative distance between $A$ and $B$ particles discussed in Appendix~\ref{appendA}. Nevertheless, the two delocalized $AB$ pairs maintain close proximity within the tetramer phase, as indicated by the local maxima in the intraspecies correlators [Fig.~\ref{pdiag2d}(b2)].

However, as $|U_{AB}/U|$ increases, the interspecies separation decreases more rapidly for particles belonging to the same dimer than for those in different dimers. Consequently, each pair becomes highly localized, leading to a separation of the two dimers. This transition is evident from the sharp increase in the on-site interspecies correlator and the concurrent decrease in the intraspecies correlator at nearest-neighbor sites in the regime $ -0.75 \lesssim U_{AB}/U \lesssim -0.42$, where the system consists of two separated dimers [Figs.~\ref{pdiag2d}(b1) and \ref{pdiag2d}(b2)]. 
As $|U_{AB}/U|$ increases further, the distance between these localized dimers continues to decrease. Eventually, for $U_{AB}/U < -0.75$, the dimers preferentially occupy a common neighborhood to form a strongly bound tetramer, contributing to $E_b < 0$. Therefore, the 2D lattice structure enhances the stability of the weakly bound tetramer regime, giving rise to a characteristic narrow corridor of $E_b < 0$ at lower $|U_{AB}/U|$. Understanding the physical mechanism underlying this regime and the use of two-body correlators to distinguish these delocalized tetramers from the strongly bound structures typically observed at larger $|U_{AB}/U|$ constitute a key highlight of this work.

\subsubsection{Hopping anisotropy ($J_y$ tuning)} \label{tune_jy}

To further elucidate the impact of connectivity and dimensionality on the phase diagrams, we examine an $8 \times 5$ lattice with tunable interchain 
hopping $J_y$, as shown in Figs.~\ref{pdiag_anisotropy}(a1)–\ref{pdiag_anisotropy}(a4). In the decoupled limit 
($J_y = 0$), the system reduces to an array of independent 1D chains, 
and the phase diagram closely resembles the strictly 1D case [compare Figs.~\ref{pdiag_anisotropy}(a1) and \ref{pdiag2d}(a1)]. However, the underlying 
ground-state manifold is highly degenerate, and therefore, nontrivial; see, for example, Fig.~\ref{pdiag_anisotropy}(b), where we show energies of the low-lying eigenstates (indexed by label $\alpha$) as a function of $J_y$ for $U_{AB}/U = -0.6$ and $U/J_x = 10$. Due to the strong sensitivity of the entanglement entropy to different linear combinations of states within the degenerate ground-state manifold, we do not show the (typical green curve) boundary between the weakly and strongly correlated phases.

In the 1D tetramer regime, as previously discussed [see Fig.~\ref{pdiag1d}(f1)], the 
constituent dimers preferentially occupy nearest-neighbor sites within a single chain to 
maximize four-body binding. For an array of $L_y = 5$ decoupled chains ($J_y =0$), the tetramer ground state leads to a 
five-fold degenerate ground state, where a localized tetramer can form in any one of the 
five independent chains. Conversely, in the dimer regime of a single 1D lattice, the dimers maximize their spatial 
separation to minimize kinetic constraints. Therefore, when considering $L_y =5$ chains in the $J_y = 0$ limit, two dimers can 
occupy any two distinct chains, resulting in a tenfold degenerate ground-state manifold of 
$\binom{5}{2}$ configurations. Meanwhile, we have found that the first excited manifold is fivefold degenerate, 
consisting of states where both dimers are constrained to the same row, recovering the 1D-like configuration (not shown). 

The introduction of finite interchain hopping ($J_y > 0$), however, removes these degeneracies, as 
illustrated by the energy spectrum in Fig.~\ref{pdiag_anisotropy}(b). For $J_y = 0$, at 
the representative parameters $U_{AB}/U = -0.6$ and $U/J_x = 10$, the system resides in the 
ten-fold degenerate dimer manifold in the ground state; see the inset of Fig.~\ref{pdiag_anisotropy}(b). As $J_y/J_x$ increases, the tetramer configuration gains 
significant kinetic energy through transverse hopping pathways (along $L_y$), eventually becoming the ground state of the system. Consequently, 
an increase in $J_y$ drives a systematic expansion of the tetramer phase at the expense 
of the dimer phase, as evidenced in Figs.~\ref{pdiag_anisotropy}(a1)–\ref{pdiag_anisotropy}(a4). This behavior 
underscores how transverse connectivity stabilizes multiparticle bound states by 
relaxing the kinetic penalties inherent to 1D.

To elucidate the structural evolution within the tetramer phase and the dimer-to-tetramer crossover driven by $J_y$, we compute two-body correlation functions for the interaction parameters indicated by the markers in Figs.~\ref{pdiag_anisotropy}(a1)--\ref{pdiag_anisotropy}(a4). Specifically, we fix one particle at $(i', j') = (3, 2)$, corresponding to a local maximum of the one-body density in the bulk, and evaluate the correlations as the coordinates $(i, j)$ of the second particle span the lattice. Given that the $J_y=0$ ground states exhibit fivefold and tenfold degeneracies for the tetramer and dimer phases, respectively, we perform our analysis on a representative configuration for each case. For the tetramer, we select the state where all particles occupy the $j = 2$ row; for the dimer phase, we consider the configuration where the two dimers reside in the $j = 2$ and $j = 3$ rows.

We first examine the parameter set, $U_{AB}/U = -0.6$ and $U/J_x = 10$ [square markers in Figs.~\ref{pdiag_anisotropy}(a1)--\ref{pdiag_anisotropy}(a4)], where the ground state undergoes a transition from a dimer to a tetramer phase for $J_y \gtrsim 0.3 J_x$. At $J_y = 0$, the interspecies correlator is most prominent at site $(i, j) = (3, 2)$, signaling strong on-site pairing, with additional significant correlations appearing at neighboring sites along the $j = 3$ row [Fig.~\ref{pdiag_anisotropy}(c1)]. Consistent with this, the intraspecies correlator is non-zero only along the $j = 3$ row, peaking at the $(3, 3)$ site [Fig.~\ref{pdiag_anisotropy}(e1)]. As $J_y$ increases, the interspecies correlator at the reference site progressively drops while enhancing at the surrounding sites [Figs.~\ref{pdiag_anisotropy}(c2)--\ref{pdiag_anisotropy}(c4)]. Similarly, the intraspecies correlator strengthens across all neighboring sites while developing a local dip at the reference site [Figs.~\ref{pdiag_anisotropy}(e2)--\ref{pdiag_anisotropy}(e4)].

A similar trend is observed for the parameter set $U_{AB}/U = -0.85$ and $U/J_x = 8$ [triangular markers in Figs.~\ref{pdiag_anisotropy}(a1)--\ref{pdiag_anisotropy}(a4)], where the tetramer becomes increasingly bound, as evidenced by a progressively larger negative binding energy with increasing $J_y$. At $J_y = 0$, both the interspecies [Fig.~\ref{pdiag_anisotropy}(d1)] and intraspecies [Fig.~\ref{pdiag_anisotropy}(f1)] correlators strongly resemble the strictly 1D case discussed earlier. As $J_y$ increases, the correlation density is redistributed toward neighboring sites, reflecting a shift in the spatial structure; see Figs.~\ref{pdiag_anisotropy}(d2)--\ref{pdiag_anisotropy}(d4) and \ref{pdiag_anisotropy}(f2)--\ref{pdiag_anisotropy}(f4). This redistribution signifies the formation of a tightly bound tetramer in the 2D lattices where particles maintain spatial proximity while minimizing double occupancy, a consequence of the competition between increased hopping and effective on-site repulsion.

\begin{figure}[t]
   \includegraphics[width=0.48\textwidth]{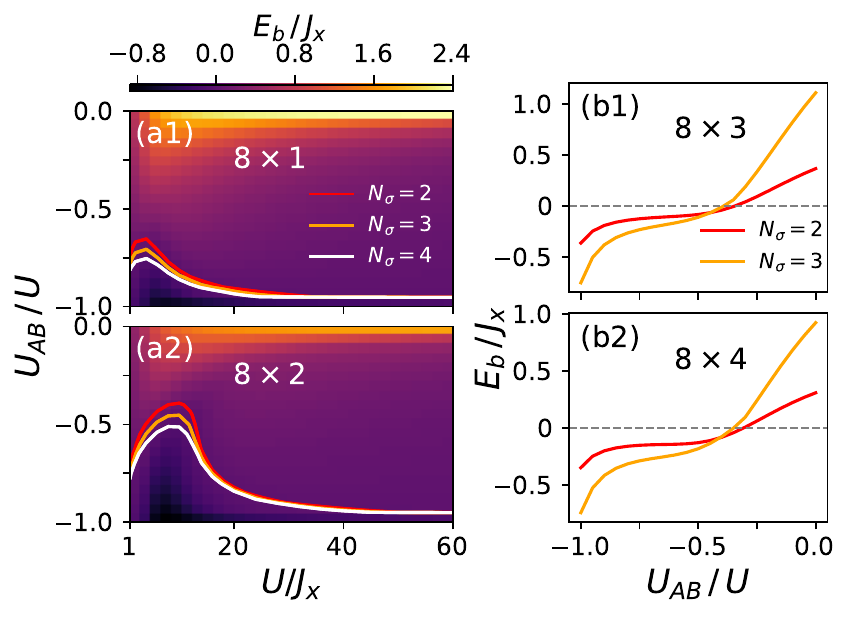}
      \caption{Bound complexes with different particle numbers. (a1),(a2) Binding energy $E_b$ as a function of $U_{AB}/U$ and $U/J_x$ for the $N_A = N_B = 3$ system.  The transition boundary separating the $N_A + N_B$ bound complexes $(E_b < 0)$ formed of $N_A=N_B=2,3,4$ is displayed with red, orange, and white lines, respectively, for $8\times1$ and $8\times2$ lattices. (b1),(b2) $E_b$ as a function of $U_{AB}/U$ at fixed interaction strength $U/J_x=10$ for $8\times3$ and $8\times4$ lattices. The region with the negative binding energy shrinks along the $U_{AB}/U$ axis with increasing $N_{\sigma}$.
      }
  \label{pdag_hex_oct}
\end{figure}

\subsection{Bound complexes with larger particle number} \label{diff_particles}

So far, we have exemplified the role of dimensionality on few-body bound states and the associated correlation signatures by considering the $N_A = N_B = 2$ system, where the tetramer constitutes the primary bound cluster. Considering the case of isotropic hoppings ($J_y=J_x$), we now extend this discussion to higher-order bound complexes that arise for $N_A = N_B = 3$ and $4$, corresponding to hexamer ($N = 6$) and octamer ($N = 8$) states, respectively. The binding energies of the hexamer and octamer complexes are defined as
\begin{eqnarray}
    E_b^{\rm hex} &=& E_{AAABBB} - \min\{3E_{AB},\, E_{AABB} + E_{AB}\}, \nonumber \\
    E_b^{\rm oct} &=& E_{AAAABBBB} \nonumber \\
                  && - \min\{4E_{AB},\, 2E_{AABB},\, E_{AAABBB} + E_{AB}\}.
\end{eqnarray}

In Figs.~\ref{pdag_hex_oct}(a1)--\ref{pdag_hex_oct}(a2), we plot the binding energy of the $N_A = N_B = 3$ system and show the phase boundaries for $N_A = N_B = 2, 3, 4$ in a 1D chain with $L = 8$ sites, shown in panel (a1), and compare them with those for the $8 \times 2$ lattice shown in panel (a2). The region of negative binding energy ($E_b < 0$) for these complexes lies within the area enclosed by the solid curves, with each color denoting a specific particle number. As the particle number increases, the $E_b < 0$ region shrinks in the $U_{AB}/U$ parameter space. However, for large values of $|U_{AB}/U|$ and $U/J_x$, increasing the particle number results in only negligible shifts of the transition boundary between $E_b < 0$ and $E_b > 0$. This consistency persists across higher dimensions, such as the $8 \times 3$ and $8 \times 4$ lattices [Fig.~\ref{pdag_hex_oct}(b1)--\ref{pdag_hex_oct}(b2)]. Due to the rapid growth of the Hilbert space for these systems, we compute the binding energy at a fixed value of $U/J_x = 10$. We observe that for larger $N$, the onset of the bound phase ($E_b < 0$) shifts toward smaller values of $U_{AB}/U$. Most importantly, the central observation of this work, that the stability of these bound complexes is significantly enhanced when moving from 1D chains to coupled 1D arrays, holds irrespective of the total particle number $N$.

Having established that the equilibrium stability of these complexes is robust across different particle numbers and geometries, we next examine their formation dynamics. For the remainder of this work, we focus on the $N_A = N_B = 2$ case as a representative model. This choice allows for a detailed investigation of the dynamical evolution and entanglement signatures during the dimensional crossover without the prohibitive computational cost associated with larger $N$ in two-dimensional geometries.

\section{Dynamical formation of few-body bound states}\label{Dynamics}

\begin{figure}[t]
   \includegraphics[width=0.48\textwidth]{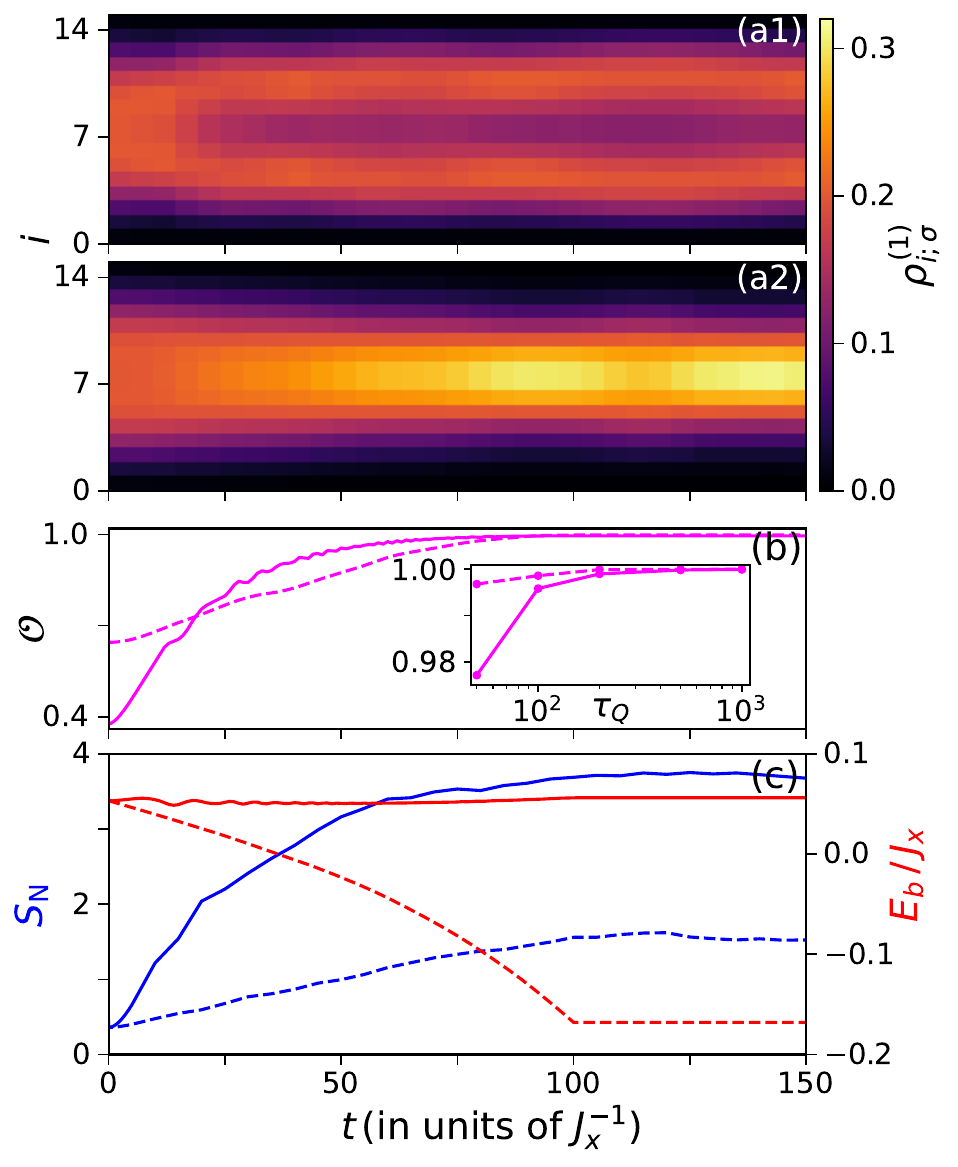}
   \caption{Dynamics following interaction quenches in a 1D lattice. (a1), (a2) Time evolution of the one-body density, $\rho^{(1)}_{i;\sigma}$, for (a1) an intraspecies interaction quench from $U=2J_x$ to $U=25J_x$ at $U_{AB} = -0.35U$, representing a weakly correlated to strongly correlated dimer transition, and (a2) an interspecies interaction quench from $U_{AB} = -0.35U$ to $U_{AB} = -0.95U$ at $U=2J_x$, representing a weakly correlated dimer to tetramer transition, for $\tau_Q=100$ . 
   (b) Temporal evolution of the fidelity $\mathcal{O}(t)$. The inset shows the fidelity at the 
   end of the quench $\mathcal{O}(t=\tau_Q)$ as a function of $\tau_Q$.  (c) Time evolution of the interspecies entanglement entropy, $S_{\mathrm{N}}$ (left axis), and the binding energy, $E_b$ (right axis). Solid and dashed lines correspond to the weakly correlated dimer to strongly correlated dimer and tetramer quenches, respectively.}
   \label{dyn_1body_den}
\end{figure}

Following our analysis of the equilibrium phase diagrams, we investigate the dynamical formation of the few-body structures. The dynamics are initiated using two protocols: (1) an interaction quench in a single 1D or 2D lattice, driving the system from a weakly correlated dimer into a strongly correlated dimer or tetramer regime; and (2) a dimensional crossover in an $8 \times 5$ lattice, induced by quenching the interchain hopping $J_y$. 

In both cases, the transition is implemented via a linear time-dependent ramp of the control parameter $\mathcal{V}(t)$, which represents the intraspecies interaction $U/J_x$, the interspecies ratio $U_{AB}/U$, or the hopping parameter $J_y$. The protocol is defined as
\begin{equation}
\mathcal{V}(t) = \mathcal{V}_{\rm ini} + \left( \mathcal{V}_{\rm fin} - \mathcal{V}_{\rm ini} \right) \frac{t}{\tau_Q},
\end{equation}
where $\tau_Q$ denotes the quench duration. For $t > \tau_Q$, the system evolves unitarily at the final parameter value $\mathcal{V}_{\rm fin}$. The resulting nonequilibrium states are characterized by monitoring the time evolution of single- and two-particle density distributions, as well as global observables such as the state fidelity, interspecies entanglement entropy, and the instantaneous binding energy.

\subsection{Interaction quench dynamics}\label{sec_dyn_intquench}

We first examine the dynamical transition from an initial weakly correlated dimer to strongly correlated dimer and tetramer phases in a 1D lattice. The system is initialized in a weakly correlated dimer phase characterized by $U = 2J_x$ and $U_{AB} = -0.35U$. We then quench the interaction strength to the strongly correlated dimer regime ($U = 25J_x$) while keeping the ratio $U_{AB}/U$ fixed  by linearly ramping $U/J_x$ over a duration $\tau_Q = 100$ (expressed in units of $J_x^{-1}$). To identify the structural reorganization in real space, we show the time evolution of the one-body density $\rho^{(1)}_{i;\sigma}$ in Fig.~\ref{dyn_1body_den}(a1). The initially delocalized density begins to reorganize around $t \approx 15$; by the end of the interaction ramp ($t=100$), a stable double-peak structure emerges. This spatial bifurcation signals the formation of two localized pairs, characteristic of the strongly correlated dimer phase. 

In contrast, a quench to the tetramer regime, characterized by $U = 2J_x$ and $U_{AB} = -0.95U$, leads to a qualitatively different density evolution [Fig.~\ref{dyn_1body_den}(a2)]. For $t \gtrsim 25$, the corresponding $\rho^{(1)}_{i;\sigma}$ progressively localizes at the center of the lattice, signaling the formation of a tightly bound tetramer.

\begin{figure}[t]
   \includegraphics[width=0.49\textwidth]{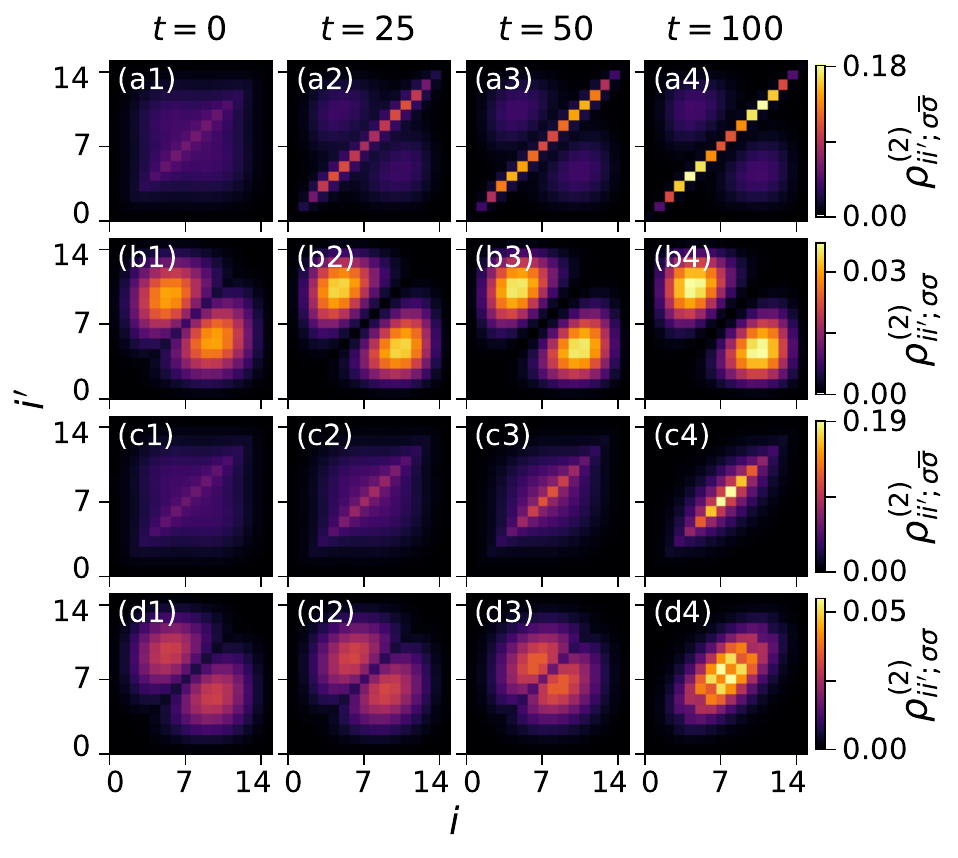}
   \caption{Evolution of two-body correlations following interaction quenches. Instantaneous profiles of the two-body reduced density matrices for (a1)--(a4), (c1)--(c4) interspecies pairs $\rho^{(2)}_{ii';\sigma\bar{\sigma}}$ and (b1)--(b4), (d1)--(d4) intraspecies pairs $\rho^{(2)}_{ii';\sigma\sigma}$ ($\sigma = A, B$) . Panels (a1)–(a4) and (b1)–(b4) depict the dynamical formation of strongly correlated dimers following a quench from $U/J_x = 2$ to $U/J_x = 25$ at $U_{AB} = -0.35U$. Panels (c1)–(c4) and (d1)–(d4) illustrate the formation of a tetramer state following a quench from $U_{AB} = -0.35U$ to $U_{AB} = -0.95U$ at $U=2J_x$. Both processes originate from a weakly correlated dimer. The diagonal and off-diagonal features track the emergence of localized spatial correlations and the development of bound-state structures over time.}
   \label{dyn_2body_corr}
\end{figure}

To further resolve the spatial rearrangement of particles, we examine the time evolution of the two-body reduced density matrices, $\rho^{(2)}_{ii';\sigma\sigma'}$, shown in Fig.~\ref{dyn_2body_corr}.
For the weakly  to strongly correlated dimer quench, the interspecies correlator $\rho^{(2)}_{ii';AB}$ [Figs.~\ref{dyn_2body_corr}(a1)--\ref{dyn_2body_corr}(a4)] develops a pronounced diagonal ridge with two symmetric maxima about the lattice center, indicating robust local $AB$ pairing within dimers. Simultaneously, the intraspecies correlator $\rho^{(2)}_{ii';\sigma\sigma}$ [Figs.~\ref{dyn_2body_corr}(b1)--\ref{dyn_2body_corr}(b4)] evolves from an initially weak diagonal depletion into a structure featuring a pronounced suppression along the diagonal ($i=i'$) accompanied by distinct off-diagonal lobes. This behavior reflects the spatial separation of same-species particles into different dimers to minimize intraspecies repulsion. Conversely, for the weakly correlated dimer to tetramer quench, $\rho^{(2)}_{ii';AB}$ [Figs.~\ref{dyn_2body_corr}(c1)--\ref{dyn_2body_corr}(c4)] becomes sharply localized at the center of the diagonal, signaling the formation of a tetramer. The corresponding intraspecies correlator [Figs.~\ref{dyn_2body_corr}(d1)--\ref{dyn_2body_corr}(d4)] exhibits only a weak diagonal depletion due to a stronger localization tendency.

To move beyond the qualitative spatial patterns observed in the density evolution and provide a quantitative assessment of the quench dynamics, we analyze several global observables. We first assess the adiabaticity of the process, the fidelity of the time-evolved state by computing the many-body overlap $\mathcal{O}(t)$, defined in Eq.~(\ref{olap})
 As shown in Fig.~\ref{dyn_1body_den}(b), the overlap for both the weakly correlated dimer to strongly correlated dimer (solid magenta line) and weakly correlated dimer to tetramer (dashed magenta line) quenches approaches unity as the dynamical evolution reaches the end of the ramp ($t \to \tau_Q$). This high fidelity demonstrates that the linear ramp is sufficiently slow to avoid significant excitations, ensuring that the dynamically generated clusters faithfully realize the intended ground-state phases; see the inset of Fig.~\ref{dyn_1body_den}(b).  The internal correlation structure of the emerging phases is further characterized by the evolution of the von Neumann entanglement entropy, $S_{\rm N}$, and the binding energy, $E_b$, shown in Fig.~\ref{dyn_1body_den}(c). Starting from a weakly entangled state, $S_{\rm N}$ increases rapidly as the interaction ramp induces strong correlations. Notably, the strongly correlated dimer phase (solid blue line) saturates at a significantly higher entanglement entropy than the tetramer phase (dashed blue line), for the chosen parameters. Concurrently, the binding energy $E_b$ (red dashed curve) evolves toward negative values (for $t \gtrsim 25$), signaling the energetic stabilization of the bound states. Therefore, these global diagnostics complement the real-space information provided by the one- and two-body density profiles. The stationary values attained by these observables at long times indicate that the quenched state closely approximates an eigenstate of the final Hamiltonian.
 
\begin{figure}[t]
   \includegraphics[width=0.48\textwidth]{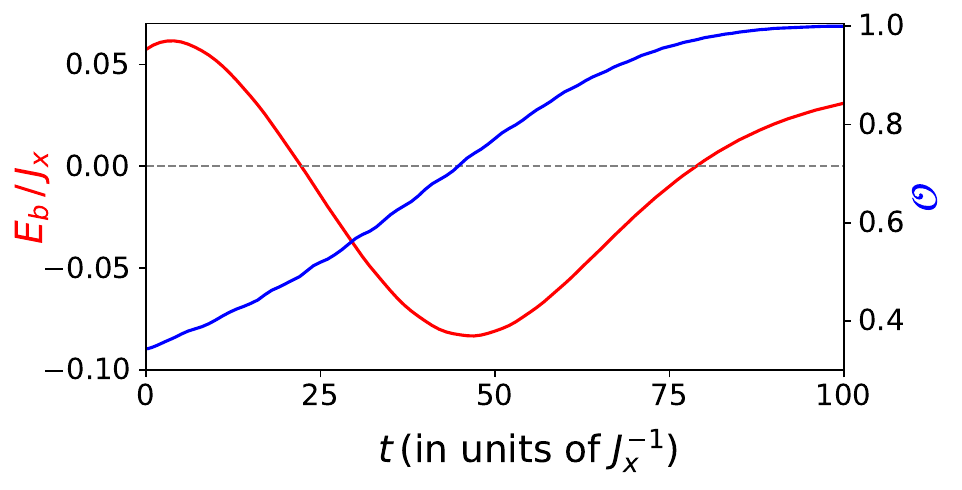}
      \caption{Weakly correlated dimer to tetramer to strongly correlated dimer quench. For an isotropic $8\times 5$ lattice, quenching $U/J_x = 2$ to $25$ at constant $U_{AB}/U = -0.6$  and $\tau_Q=100$ triggers a crossover between a weakly correlated dimer to a tetramer and finally a strongly correlated dimer phase. The tetramer regime is identified with negative $E_b$ (red curve).
      }
  \label{2d_int_qnch}
\end{figure}

Finally, for completeness, we extend this analysis to a 2D $8\times 5$ lattice with isotropic hopping. Following a quench from $U/J_x=2$ to $25$ at $U_{AB}/U = -0.6$, the evolution of $E_b$ is presented in Fig.~\ref{2d_int_qnch}  for $\tau_Q=100$. From an initial weakly correlated dimer state ($E_b > 0$), the system enters a transient tetramer regime ($E_b < 0$), reaching a minimum that reflects a strongly bound state. This feature eventually diminishes as the state dynamically evolves toward the strongly correlated dimer phase near the end of the quench. As in the 1D case, adiabaticity is preserved ($\mathcal{O} \approx 1$ at the end of quench) for the 2D transition.

\begin{figure*}[t]
   \includegraphics[width=0.98\textwidth]{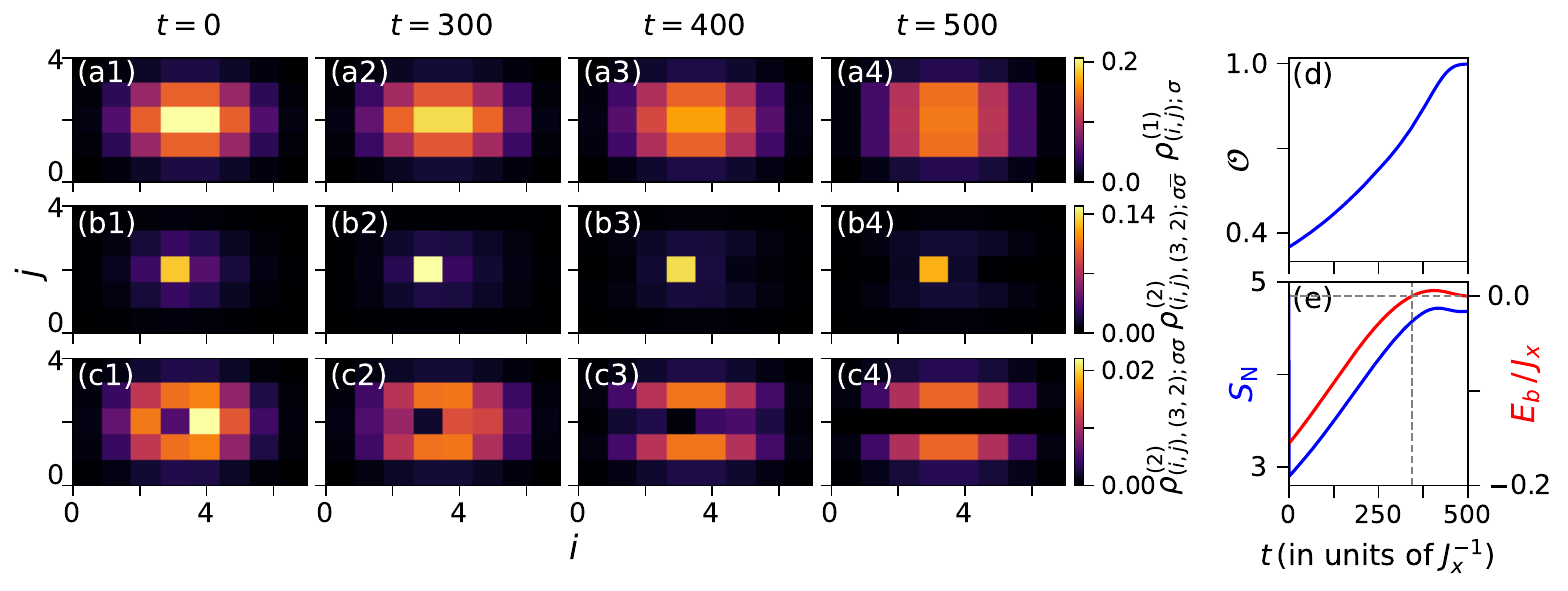}
        \caption{Geometric quench from $8\times5$ to array of uncoupled 1D chains. Dynamical crossover from an initial tetramer to a strongly correlated dimer state, triggered by linearly ramping hopping strength $J_y/J_x$ from $1$ at $t =0$ to $0$ at $t=500$ for $U/J_x=10$, $U_{AB}/U=-0.6$,  and $\tau_Q=500$. Instantaneous profiles of one-body density matrix [panels (a1)-(a4)] and  two-body  reduced density matrices showing interspecies [panles (b1)–(b4)] and intraspecies [panels (c1)-(c4)] correlations during the dynamics.(d) Overlap of the dynamical state with the equilibrium tenfold degenerate ground state manifold at $J_y = 0$, 
        achieving a value $\mathcal{O} \approx 1$ at the end of quench. (e) Entanglement entropy
        and binding energy during dynamics. The crossover from bounded state to unbounded state
        occurs at $t\approx 344$.}
  \label{dim_cross}
\end{figure*}

\subsection{Dimensional crossover between dimers and tetramers} \label{dimen_cross}

To further explore the interplay between lattice geometry and few-body complexes such as the dimer and tetramer states studied here, we investigate the dynamical crossover induced by a sudden quench of the interchain hopping parameter, $J_y$. As established in Sec.~\ref{Array1D}, reducing the interchain hopping $J_y$
  relative to the intrachain hopping $J_x$
  shrinks the tetramer region of the phase diagram [see Figs.~\ref{pdiag_anisotropy}(a1)--\ref{pdiag_anisotropy}(a4)]. Consequently, for a fixed set of interaction parameters, tuning the hopping anisotropy can trigger a transition from a 2D tetramer to a collection of 1D strongly correlated dimers.

We model this dimensional crossover using an $8 \times 5$ lattice with interaction parameters $U_{AB}/U = -0.6$ and $U/J_x = 10$. The system is initialized in the ground state of the isotropic lattice ($J_y = J_x$), where the particles form a tightly bound tetramer. We then linearly ramp the transverse hopping $J_y$ to zero over a quench duration $\tau_Q = 500$, thereby effectively decoupling the system into an array of independent 1D chains. The slow ramp is deliberately chosen to ensure adiabatic evolution as the system approaches the highly degenerate ground-state manifold of the $J_y = 0$ limit [see Fig.~\ref{pdiag_anisotropy}(b)].

The structural reorganization during this process is captured by the instantaneous one-body density, $\rho^{(1)}_{(i,j);\sigma}$, shown in Figs.~\ref{dim_cross}(a1)–-\ref{dim_cross}(a4). Initially, the density is sharply localized at the center of the lattice, a hallmark of the 2D tetramer. As $J_y$ decreases, the tetramer one-body density spreads along $L_y$. Upon reaching the decoupled limit ($J_y \to 0$), the density redistributes into a dimer phase, characterized by a nearly uniform distribution along $L_y$ within the $L_x$, except for minor edge effects.

To resolve the internal structure of these composites, we examine the two-body reduced density matrices by fixing one particle at the lattice center $(i',j') = (3,2)$. The interspecies correlator [Figs.~\ref{dim_cross}(b1)–\ref{dim_cross}(b4)] remains centered at the reference site $(i, j) = (3, 2)$ throughout the time evolution. This confirms that, even as the tetramer dissolves, strong local $AB$ pairing is preserved within the resulting dimers.
 In contrast, the intraspecies correlator 
[Figs.~\ref{dim_cross}(c1)–\ref{dim_cross}(c4)] undergoes a dramatic transformation during the quench. The initial distribution is sharply peaked at $(i, j) = (4, 2)$, indicating that the probability of finding particles of the same species at nearest-neighbor sites is maximized due to the effective dimer-dimer attraction within the tetramer, as argued earlier. As the quench progresses, this characteristic tetramer signature evolves into a structure featuring a distinct central minimum along the $L_x$ direction. By the end of the 
quench ($t=500$), the two $A$ particles (and similarly the two $B$ particles) preferentially occupy distinct 1D chains at different $L_y$ positions, thereby minimizing their mutual intraspecies repulsion $U$ in the decoupled limit.

This spatial redistribution explains the absence of the localized "humplike" features in the 2D one-body density, which are a hallmark of 1D bound complexes [compare Figs.~\ref{dim_cross}(a3) and \ref{dim_cross}(a4) with Fig.~\ref{dyn_1body_den}(a1)]. By 
populating distinct 1D chains, the particles effectively bypass the energy penalty imposed by $U$. Furthermore, this evolution confirms that the system has successfully transitioned into the ten-fold degenerate ground-state manifold of the decoupled 
limit. In this regime, the two $AB$ dimers are free to occupy any combination of the available chains to minimize the total energy, consistent with the degeneracy arguments presented in Sec.~\ref{Array1D}.

We quantify the transition's progress using the many-body fidelity $\mathcal{O}(t)$, defined as $\mathcal{O} (t) = \sum_{\alpha} |\langle \Psi^{\rm eq}_{\rm gs, \alpha} (J_y = 0) | \Psi(t)\rangle|^2$, where $\alpha$ labels the different ground states with tenfold degeneracy at $J_y = 0$. As shown in Fig.~\ref{dim_cross}(d), the fidelity starts at a low value ($\approx 0.35$), reflecting the distinct symmetry of the 2D tetramer, but converges to $O \approx 0.998$ by the end of the ramp. This confirms that the quench effectively prepares the system in the target dimer manifold. 
We also track the binding energy [right axis, Fig.~\ref{dim_cross}(e)] to assess the system's transition. The transition is marked by the vanishing of the binding energy at $t \approx 344$, signaling the point where the four-body bound state becomes energetically unstable and dissociates into two independent dimers.  The entanglement entropy grows linearly as the system explores the expanding configurational space during the crossover [Fig.~\ref{dim_cross}(e)].

\section{Conclusions}\label{sec_conclusions}

In this work, we have characterized the formation of few-body bound states, specifically dimers and tetramers, within the ground state and nonequilibrium dynamics of a particle-balanced, four-body binary mixture in an optical lattice. Using binding energy and interspecies entanglement entropy as primary diagnostics, we mapped the regions of existence of these complexes across a phase diagram spanned by the attractive interspecies and repulsive intraspecies interactions. Starting from 1D lattices, we have systematically extended our analysis to arrays of coupled chains, thereby understanding the role of dimensionality through controlled variation of the number of chains and the interchain hopping strengths. Guided by the resulting equilibrium phase diagrams, we have investigated the dynamical formation of such states following quenches of both interactions and interchain tunneling. To interpret these processes, we integrated local density matrices with global measures, such as state fidelity and entanglement growth, to quantify the efficiency of the dynamical preparation and the buildup of many-body correlations.

Focusing on the ground-state properties, we have found that as the system becomes increasingly extended in the transverse direction, the tetramer phase expands substantially, occupying regions of the phase diagram where it is absent in a strictly 1D lattice. Most importantly, by tuning the interchain hopping $J_y$ from zero to finite values in a 2D lattice, we have demonstrated that the system undergoes a transition from a highly degenerate manifold of spatially separated dimers to a localized tetramer ground state. This transition is driven by the lifting of 1D configurational degeneracies and a significant gain in kinetic energy as particles exploit transverse hopping pathways.

Microscopically, these states are uniquely identified by their distinct correlation signatures. The tetramer is characterized by a sharply peaked diagonal in the interspecies density matrix, accompanied by a minimized depletion in the intraspecies matrix. This indicates that particles of the same species preferentially occupy nearest-neighbor sites, signaling the formation of a tightly bound, four-body cluster.  In contrast, the strongly correlated dimer phase is distinguished by two spatially separated peaks in the one-body density and the emergence of symmetric, off-diagonal lobes in the intraspecies correlations. These lobes reflect the spatial separation of two distinct composite pairs, driven by effective intraspecies repulsion. As the effective dimensionality of the system increases, either by adding additional lattice chains or by tuning the interchain hopping parameter, the nearest-neighbor distribution of intraspecies two-body correlations becomes increasingly concentrated around the reference site, while preserving a dip at the reference-site position. This indicates that, compared to a strictly 1D lattice, a 2D configuration relaxes kinetic constraints, allowing proximity between particles and thereby enhancing the likelihood of bound-state formation. In addition, by extending our analysis to larger particle numbers, we identify the formation of higher-order bound complexes, including hexamers and octamers. Although the parameter regime supporting bound states becomes progressively narrower with increasing 
$N$, the enhancement of stability with increasing dimensionality remains robust. This demonstrates that the increased dimensionality provides a generic mechanism for stabilizing composite bound states beyond the few-body limit.

Our dynamical simulations demonstrate that strongly correlated dimer and tetramer states can be prepared from a weakly correlated dimer via experimentally viable interaction and geometric quenches. Following linear ramps of the interaction strength, the system undergoes a distinct structural bifurcation depending on the final interaction regime. In the weakly to strongly correlated dimer quench, the initially delocalized density evolves into a stable double-peak structure, characterized microscopically by off-diagonal lobes in the intraspecies two-body density matrices, signaling the spatial separation of repulsive pairs. Conversely, the weakly correlated dimer to tetramer quench triggers the formation of a single sharply localized one-body density peak at the lattice center, supported by a strong interspecies two-body diagonal correlations. The high state fidelities $\mathcal{O} \rightarrow 1$  and the saturation of binding energy at the characterized values confirm that these quenches faithfully prepare the target ground-state phases. By quenching the interchain hopping $J_y$, we realize a controlled transition from a two-dimensional tetramer to a collection of one-dimensional dimers. This geometric quench drives a transverse delocalization of the one-body density and a reorganization of two-body correlations: While local $AB$ pairing remains robust, particles of the same species redistribute across distinct chains to minimize intraspecies repulsion. The transition of the binding energy to $E_b > 0$ and the convergence of the fidelity to the ten-fold degenerate manifold of the decoupled limit confirm the dynamical dissociation of the four-body cluster into spatially separated dimers.

Looking forward, the insights gained in this work provide a robust foundation for several future research avenues. A natural extension is to study the  hopping imbalance case \cite{giri_2021, giri_2022}, or species imbalance case, and how these few-body bound complexes gradually evolve into self-bound lattice liquids at finite density~\cite{machida_2022, morera_2023, perezcruz_2025}. Addressing this crossover will likely require a synergistic approach that combines exact diagonalization with advanced \emph{ab initio} methods such as ML-MCTDHX \cite{Cao2013,cao2017, Englezos2024,paolo_2025}. A particularly intriguing direction involves extending this framework to Bose-Fermi or Fermi-Fermi mixtures \cite{onofrio_2016}, where Pauli blocking is expected to introduce unique constraints on the formation of bound states. Furthermore, investigating the role of long-range interactions, such as dipolar couplings \cite{chomaz_2022}, could reveal additional features due to anisotropic interactions (depending on the orientation of dipoles) in the few-body phase diagram. Finally, our established quench protocols offer a clear roadmap for future experiments. Utilizing quantum gas microscopy \cite{gross_2017,gross_2021}, it is now feasible to probe the real-time synthesis and fragmentation of these exotic quantum complexes with single-site resolution, providing a direct experimental test of the correlation dynamics predicted here.

{\it Acknowledgments}--- D.G. acknowledges discussions with Ralf Klemt at a conference at MPIPKS, Dresden. We also thank T.~Arnone~Cardinale, L.~Chergui, M.~Schubert,  and P.~St\"urmer for discussions. Financial support from the Knut and Alice Wallenberg Foundation through the Wallenberg Centre for Quantum Technology as well as Grant~KAW2023.0322, and the Swedish Research Council (Grant No.~2022-03654 VR) is gratefully acknowledged.  Part of the computations were enabled by resources provided by the National Academic Infrastructure for Supercomputing in Sweden (NAISS), partially funded by the Swedish Research Council through Grant Agreement No. 2022-06725 VR. K.M. acknowledges financial support from the JSPS Postdoctoral Fellowship (Fellowship No.~P25029) and
JSPS KAKENHI Grant (Grant No. JP26K00638).

\emph{Data availability}--- The data that support the findings of this article are openly available at \cite{gaur_2026_zenodo}.

\appendix


\section{Two-body states-- $AB$ pair} \label{appendA}

\begin{figure}[t]
   \includegraphics[width=0.48\textwidth]{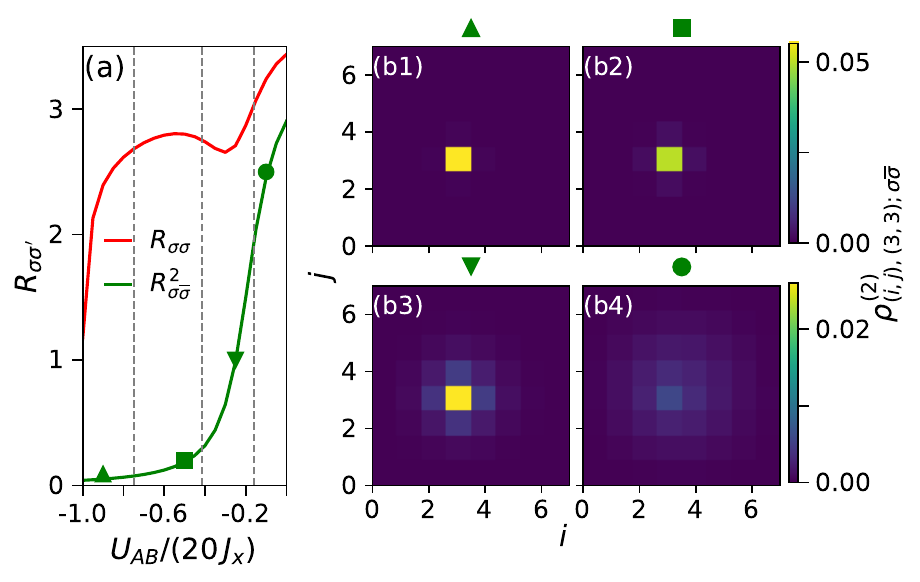}
      \caption{{Two-body $AB$ pair on a $8\times 8$ lattice. (a) Size of the $AB$ pair, $R^2_{\sigma \overline{\sigma}}$ (green curve), for a two-particle system as a function of $U_{AB}/(20\,J_x)$. Corresponding average separation between two particles of the same species $R_{\sigma \sigma}$ (red curve), for a four-particle system comprising two particles of each species as a function of $U_{AB}/U$ at $U/J_x=20$. (b1)-(b4) Two-body reduced density matrices at $U_{AB}/(20\,J_x)= -0.9, -0.5, -0.25$, and $-0.1$, indicated by the green markers in panel (a). }
      }
  \label{1p1_dimer}
\end{figure}

With two particles, one of each species, on an $8\times8$ lattice with isotropic hoppings, the ground state corresponds to a two-particle bound-state, an ``$AB$ pair," arising from the attractive interspecies interaction, which lowers the total energy by inducing pairing between the two species. However, depending on the strength of the attractive interaction, the pair has a sharp or delocalized spatial extent. These two-body pairs serve as the building blocks of larger bound complexes, and define the geometric structure there. Thus, it is meaningful to quantify the spatial extent of the pair by computing the separation between its two constituents, defined as
\begin{equation}
    R_{\sigma\sigma^{\prime}} = a\sum_{(i,j),(i',j')} \rho^{(2)}_{(i,j),(i',j');\,\sigma\sigma'}
                 \sqrt{((i-i')^2 + (j-j')^2)},
\end{equation}
where $a$ denotes the lattice spacing.
Figure~\ref{1p1_dimer}(a) shows $R^2_{\sigma\overline{\sigma}}$ (green curve), the size of the $AB$ pair as a function of $U_{AB}/(20J_x)$, clearly distinguishing regimes of tightly bound and weakly bound $AB$ pair. 
Figures~\ref{1p1_dimer}(b1)-\ref{1p1_dimer}(b4) present the two-body reduced density matrices for selected values of $U_{AB}/J_x$, indicated by green diamonds in panel (a). For large $|U_{AB}/J_x|$, the correlator is sharply localized at the reference site, with only minimal spreading to neighboring sites [panels (b1) and (b2)], characteristic of a strongly localized $AB$ pair. In contrast, for smaller $|U_{AB}/J_x|$, the correlator becomes increasingly delocalized [panles (b3) and (b4)], consistent with a larger spatial extent of the $AB$ pair. 
Finally, returning back to the four-particle case discussed in the main text, we calculate the average separation between particles of the same species. In this case, we choose on-site intraspecies interaction $U=20\,J_x$, and vary the ratio $U_{AB}/U$ to facilitate meaningful connections with the two-particle case. The intraspecies separation ($R_{\sigma\sigma}$) is shown by the red curve in Fig.~\ref{1p1_dimer}(a).
 The gray dashed line marks the phase boundary separating distinct four-particle phases, as shown in Fig.~\ref{pdiag2d}(a4). For $-1<U_{AB}/U<-0.75$, the tetramer is formed from sharply localized $AB$ pairs. As $|U_{AB}/U|$ decreases, the $AB$ pair becomes increasingly delocalized, and the system eventually transitions into a weakly correlated dimer phase at sufficiently small values of $|U_{AB}/U|$, as evidenced by $R^2_{\sigma\overline{\sigma}} \sim R_{\sigma\sigma}$. Prior to this transition, however, diffused $AB$ pairs can still bind to form a tetramer. This occurs in the range $-0.42<U_{AB}/U<-0.16$, where $R_{\sigma\sigma}$ (red curve) exhibits a local minimum, consistent with the formation of a bound state.

\begin{figure}[t]
   \includegraphics[width=0.48\textwidth]{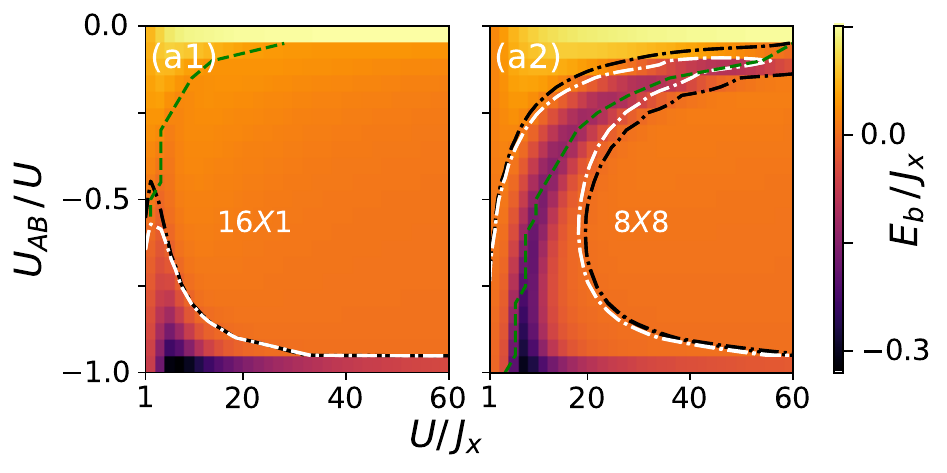}
      \caption{Effect of boundary conditions. Binding energy for the 1D and 2D lattices with PBCs. Dash-dotted white (black) curve displays the boundary separating the tetramer phase for $N=4$ in 1D and 2D lattices with open (periodic) boundary conditions. The green dashed curve, which demarcates the weak and strongly correlated phases, remains similar for the OBC and PBC cases.
      }
  \label{pbc_pdiag}
\end{figure}

\section{Effect of boundary conditions} \label{appendB}

To assess the role of finite-size effects in our study, we present results obtained using PBCs in all spatial directions. For this analysis, we consider the case $N_A = N_B = 2$ on a 1D chain with $16$ sites and on a 2D $8 \times 8$ lattice, allowing us to examine the different phases of the system. Figure~\ref{pbc_pdiag} shows the binding energy, where the black dash-dotted curve marks the boundary separating the tetramer and dimer phases. For comparison, the corresponding transition boundary obtained with OBC is shown by the white dash-dotted curve. We observe that the tetramer phase is slightly enlarged under PBCs, extending toward lower values of $|U_{AB}/U|$. However, the overall structure of the phase diagram remains qualitatively unchanged for the two choices of boundary conditions. Most importantly, we find that, compared to the 1D case, the tetramer phase occupies a significantly larger region of the phase diagram in the 2D lattice. This clearly indicates that the stabilization of bound states is enhanced with increasing lattice dimensionality. The green dashed curve, which demarcates the weak and strongly correlated phases, remains similar for the OBC and PBC cases.

\bibliography{ref}{}

\end{document}